# Computer-aided diagnosis system for Alzheimer's disease using principal component analysis and machine learning-based approaches

Lilia Lazli

Department of Computer and Software Engineering, Polytechnique Montréal,  University of Montreal, 2500 Chem. de Polytechnique, Montreal, Quebec H3T 1J4, Canada

lilia.lazli@polymtl.ca

**Abstract.** Alzheimer's disease is a severe neurological brain disorder. It is not curable, but earlier detection can help improve symptoms in a great deal. The machine learning-based approaches are popular and well-motivated models for many medical image processing tasks such as computer-aided diagnosis. These techniques can vastly improve the process for accurate diagnosis of Alzheimer's disease. In this paper, we investigate the performance of these techniques for Alzheimer's disease detection and classification using brain MRI and PET images from the OASIS database. The proposed system takes advantage of the powerful artificial neural network and support vector machines as classifiers, as well as principal component analysis as a feature extraction technique. The results indicate that the combined scheme achieves good accuracy and offers a significant advantage over the other approaches.

**Keywords**: Alzheimer's disease, Computer-aided diagnosis system, Principal component analysis, Artificial neural network, Support vector machines, MRI and PET OASIS images.

## 1  INTRODUCTION

Alzheimer's disease (AD) is a progressive degenerative brain disorder that gradually destroys memory, reason, judgment, language, and ultimately the ability to perform even the simplest of tasks. An automated AD classification system is crucial for the early detection of disease. This computer-aided diagnosis (CAD) system can help expert clinicians to prescribe the proper treatment and preventing brain tissue damage.

In recent years, researchers have developed several CAD systems. They have developed rule-based expert systems from the 1970s to 1990s and supervised models from 1990s [1]. Moreover, several approaches have been proposed in the literature aiming at providing an automatic tool that guides the clinician in the AD diagnosis process [2]. These approaches can be categorized into two types: univariate approaches including the statistical parametric mapping (SPM) and multivariate approaches such as the voxels-as-features (VAF) approach [2]. Despite the efforts of researchers, developing an automated AD classification model remains a rather challenging task. From previous research in the medical domain, it has been proved that Magnetic Resonance Imaging (MRI) and Positron Emission Tomography (PET) scans can perform a significant role for early detection of AD [3]. For our research work, we analyze this type of data using machine learning based model for AD classification.

The machine learning has shown a prominent result for organ and substructure segmentation, several diseases classification in areas of pathology, brain, breast, bone, retina, etc. But there is little existing work for AD detection using machine learning models. Therefore, an effective machine learning model is proposed, which will help physicians working on the diagnosis of AD and help them to prescribe prompt treatment for AD patients.

We developed a multi-modal classification model using a principal component analysis-based approach (PCA) in combination with supervised learning methods. We used PCA for feature extraction and support vector machines (SVMs) and artificial neural network (ANN) classifiers are trained on the features extracted from the neurological images, to detect AD from MRI and PET data. We demonstrate the performance of the CAD system on the Open Access Series of Imaging Studies (OASIS) database.

The rest of the paper is organized as follows: Section 2 presents our proposed CAD system which combines the advantages of both PCA, and machine learning based supervised classifiers. Section 3 presents an evaluation of these techniques with experimental details and results described. Finally, we conclude the article in Section 4 and present future work.

## 2  METHOD

A generic automated AD detection and classification framework is shown in Fig. 1. The task of the supervised learner is to predict the class of the input object (AD patient or healthy) by checking several learning examples. In this work, SVMs and ANN are trained on the PCA features extracted from the neurological images. Below is a summary description of the three approaches proposed for our CAD system.

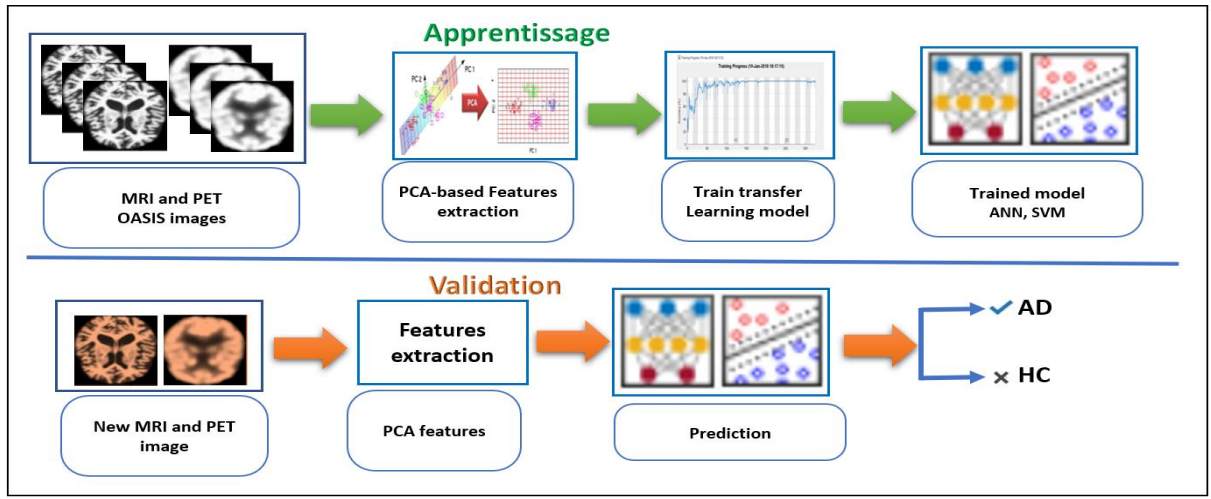

**Fig. 1**: Block diagram of a generic Alzheimer's disease CAD system.

### 2.1 Principal component analysis

PCA has been successfully applied for different image classification purposes, because it is a simple and non-parametric method of extracting relevant information from confusing data sets [4]. PCA generates an orthonormal basis vector that maximizes the scatter of all the projected samples. After the preprocessing steps, the *n* remaining voxels for each subject are rearranged into a vector form. Let $X = (x_{ij})_{n \times m}$ be the sample set of these vectors, where *n* is the number of samples (patients), $m$ is the number of features and $x_{ij}$ represents the $j^{th}$ feature value of the $i^{th}$ sample, where $i =1,2,…,n$ and $j=1,2,…,m$. In this research, some gray-scale features are selected as characteristics of the sample. The chosen gray-scale features are: mean, variance, skewness, kurtosis, energy and entropy. The specific steps of PCA are as follows:

1) Standardize the original data, all the samples are subtracted from the mean value of corresponding feature.

$$\bar{x}_j = \frac{1}{n}\sum_{i=1}^{n} x_{ij} \tag{1}$$

2) Calculate the covariance matrix $P = (r_{jk})_{m \times m}$, where $r_{jk}$ represents the correlation between the $j^{th}$ and $k^{th}$ feature.

$$P = \begin{bmatrix} r_{11} & \cdots & r_{1m} \\ \vdots & \ddots & \vdots \\ r_{m1} & \cdots & r_{mm} \end{bmatrix} \tag{2}$$

3) Calculate the eigenvalue $\lambda_i$ and the eigenvector $e_i$ of the covariance matrix $P$:

$$\lambda_i e_i = P e_i \tag{3}$$

4) Record the resulting eigenvalues in the order of large to small: $\lambda_1 \geq \lambda_2 \geq \cdots \geq \lambda_k$. Calculate the contribution rate of each principal component.

$$\frac{\lambda_g}{\sum_{g=1}^{k} \lambda_g} \tag{4}$$

The higher the contribution rate is, the stronger the information of the original variables contained in the principal component is.

5) Transform the original sample matrix $X$ into a new matrix $Y = (Y_{ij})_{n \times m_1}$, where $i = 1,2,…,n$ and $j = 1,2,…,m_1$.

$$Y = X \times [e_1, e_2, \ldots e_{m_1}] \tag{5}$$

where $[e_1, e_2, \ldots e_{m_1}]$ represents a new feature space composed of $m_1$ feature vectors. $m_1$ are the principal components extracted by PCA.

## 2.2 Support Vector Machines

SVMs are a popular tool for classification of data that is independent and identically distributed [5]. SVMs try to maximize the margin between classes (here using the simple linear feature space $x_i . x_j$), by finding the optimal $\alpha_i$ values in the following quadratic programming problem (represented in dual Lagrangian form where $C$ is a constant that bounds the misclassification error):

$$max \sum_{i=1}^{N} \alpha_i - \frac{1}{2} \sum_{i=1}^{N} \sum_{j=1}^{N} \alpha_i \alpha_j y_i y_j (x_i . x_j) \tag{6}$$

subject to: $0 \leq \alpha_i \leq C$ and $\sum_{i=1}^{N} \alpha_i y_i = 0$

Unlabelled instances are classified using the learned parameters $\alpha_i$ and bias $b$, by taking the sign of the following decision function:

$$f(x) = \sum_{i=1}^{N} \alpha_i y_i (x.x_i) + b \qquad (7)$$

## 2.3 Artificial neural network

An ANN is an information processing paradigm that is inspired by the way biological nervous systems, such as the brain and processes information [5]. ANN can be viewed as weighted directed graphs in which artificial neurons are nodes and directed edges (with weights) are connections between neuron outputs and neuron inputs. Learning process in the ANN context can be viewed as the problem of up dating network architecture and connection weights so that a network can efficiently perform a specific task.

The ability of ANN to automatically learn from examples makes them attractive and exciting. The development of the backpropagation learning algorithm for determining weights in a multilayer perceptron has made these networks the most popular among ANN researchers. For our experiments a feed-forward neural network (FFNN) was used. This type of ANN is static, that is, it produces only one set of output values rather than a sequence of values from a given input. FFNN is memory-less in the sense that his response to an input is independent of the previous network state. The following network configuration was used:

- One hidden layer and increasing number of neurons and a linear output layer.
- Hyperbolic tangent sigmoid transfer function for output layer:

$$f(n) = 2/(1 + exp(-2*n)) - 1 \qquad (8)$$

- Weight and bias values are updated according to Levenberg– Marquardt optimization.
- Gradient descent with momentum weight and bias is used as learning function.

## 3 EXPERIMENTS AND RESULTS

### 3.1 Dataset description and Image preprocessing

We have experimented the performance of the proposed model on the OASIS database [6] which provides T1-weighted MRI and PET scans with demographics and clinical assessment data. There are 300 subjects (210 AD patients and 90 healthy control subjects) aged 18 to 96, and for each of them, 3 or 4 T1-weighted MRI scans are available.

All the images were normalized through a general affine model, with 12 parameters using the SPM5 software [7]. After the affine normalization, the resulting image was registered using a more complex non-rigid spatial transformation model [7]. The data was preprocessed with an extensive MR preprocessing pipeline to reduce the effects of noise, inter-slice intensity variations, and intensity inhomogeneity.

### 3.2 Implementation Details

After the preprocessing steps, a $34 \times 47 \times 39$ voxel-sized brain representation for each subject is obtained. Thus, each subject is represented by grayscale features and is collapsed into a new feature space by applying PCA-based feature extraction.

5-fold cross validation is performed on the dataset. For each fold, we have used 70% as training data, 10% as validation data and 20% as test data. Besides the accuracy rate of the classification system (see Tab.1. for the accuracy classification results), sensitivity and specificity are the most widely statistics used to describe a diagnostic test (see Fig. 2. for the confusion matrix).

Our proposed PCA-ANN/SVM model can detect early AD and successfully classify the major AD patients and discriminate them from HC subjects. Preliminary results of evaluating the complete CAD system are shown to be more useful for separating NORMAL and AD classes. Note that PCA-SVMs accuracy was the best in all images. In any case, all the experiments increased the sensitivity–and therefore, the final accuracy rate obtained by the baseline VAF-SVMs approach.

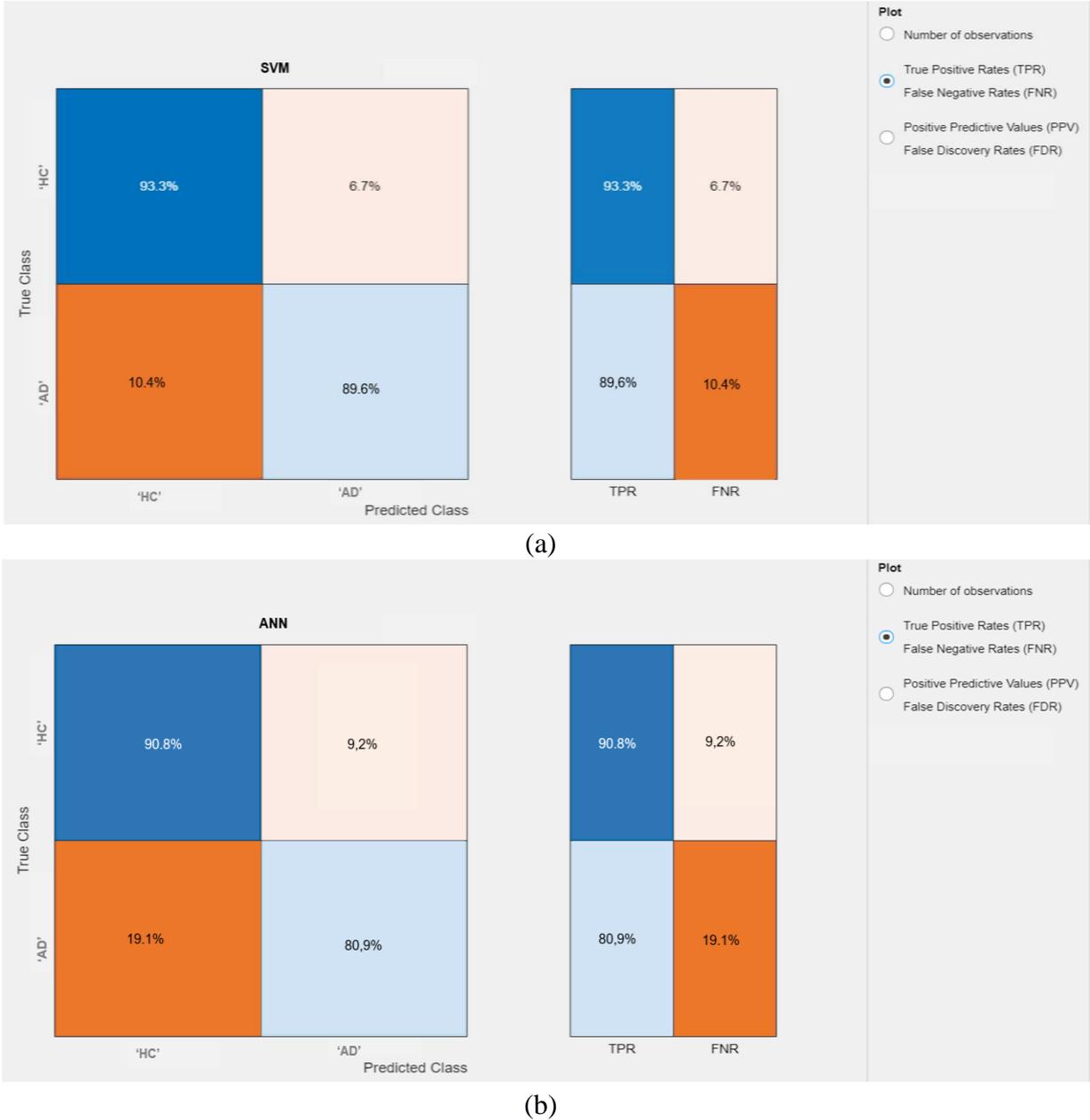

(a)

(b)

**Fig. 2**: Confusion matrix, (a) SVM, (b) ANN.

Table 1: Five-fold cross validation performance accuracy

| PCA-ANN | PCA-SVM | VAF-SVM |
|---|---|---|
| *Training results*<br>Accuracy (Validation): 71,2%<br>Total cost (Validation): NA<br>Prediction speed: ~6000 obs/sec<br>Training time: 7,7715 sec | *Training results*<br>Accuracy (Validation): **73,8%**<br>Total cost (Validation): 32<br>Prediction speed: ~2000 obs/sec<br>Training time:1,5703 sec | *Training results*<br>Accuracy (Validation): 64,4% |
| *Test results*<br>Accuracy (Test): 88,2%<br>Total cost (Validation): NA | *Test results*<br>Accuracy (Validation): **91,9%**<br>Total cost (Validation): 29 | *Test results*<br>Accuracy (Validation) : 66,3% |
| *Model type*<br>Preset: Feed-forward neural network<br>Number of fully connected layers: 1<br>First layer size: 100<br>Activation: ReLU<br>Iteration limit: 1000<br>Regularization strength: (Lambda): 0<br>Standardize: Yes | *Model type*<br>Preset: Medium Gaussian SVM<br>Kernel function: Gaussian<br>Kernel scale: 2.8<br>Box constraint level: 1<br>Multiclass method: One-vs-One<br>Standardize: True | |

## 4 CONCLUSION

Machine learning studies using neuroimaging data for developing diagnostic tools helped a lot for automated AD classification. In this work, a complete CAD system for the diagnosis of the early AD has been presented. PCA has been evaluated on the MRI/PET database as feature extraction technique. After that, the features are fed into machine learning models such as SVMs and ANN classifiers. The PCA-SVMs scheme works in general better than PCA-ANN, yielding in most cases higher accuracy rates when the same features are used. Moreover, it's more efficient than the baseline model based on VAF approach.

There are several improvements possible for the proposed CAD system. To get better performance, we hope to work with other more extensive AD datasets such as ADNI and apply other types of ANN and SVMs as well as other machine learning based classifiers. The PCA used for feature extraction, looks for the principal axis direction which is used to effectively represent the common features of the same kind of samples. This is very effective for representing the common features of the same kind of data samples, but it is not suitable to distinguish different sample classes. Therefore, PCA needs to be combined with other feature dimensionality reduction algorithms such as linear discriminant analysis to achieve the purpose of feature extraction.

**Acknowledgments**

This project was supported by the Fonds de recherche du Québec-Nature et Technologies (FRQNT) under grant # B3X-314498.

The OASIS-3 data were provided in part by OASIS Longitudinal Multimodal Neuroimaging: Principal Investigators: T. Benzinger, D. Marcus, J. Morris; NIH P30 AG066444, P50 AG00561, P30 NS09857781, P01 AG026276, P01 AG003991, R01 AG043434, UL1 TR000448, R01 EB009352. AV-45 doses were provided by Avid Radiopharmaceuticals, a wholly owned subsidiary of Eli Lilly.